\documentclass[utf8]{FrontiersinHarvard} 

\usepackage{url,hyperref,lineno,microtype,subcaption}
\usepackage[onehalfspacing]{setspace}
\usepackage{gensymb}
\usepackage{siunitx}
\usepackage{physics}
\usepackage{fixmath}
\usepackage{upgreek}
\usepackage[normalem]{ulem}

\newcommand{\vect}[1]{\boldsymbol{#1}}

\def\keyFont{\fontsize{8}{11}\helveticabold }
\def\firstAuthorLast{Zhou {et~al.}} 
\def\Authors{Zhenjun Zhou\,$^{1,2,3,*}$, Chaowei Jiang\,$^{4,*}$, Xiaoyu Yu\,$^{1}$, Yuming Wang\,$^{5}$, Yongqiang Hao\,$^{1,2}$, and Jun Cui\,$^{1,2}$}
\def\Address{$^{1}$Planetary Environmental and Astrobiological Research Laboratory (PEARL), School of Atmospheric Sciences, Sun Yat-sen University, Zhuhai, China \\
$^{2}$Key Laboratory of Tropical Atmosphere-Ocean System, Sun Yat-sen University, Ministry of Education, Zhuhai, China \\ 
$^{3}$CAS Center for Excellence in Comparative Planetology, China\\
$^{4}$Institute of Space Science and Applied Technology, Harbin Institute of Technology, Shenzhen 518055, People's Republic of China\\
$^{5}$CAS Key Laboratory of Geospace Environment, University of Science and Technology of China, Hefei, Anhui 230026, China
}
\def\corrAuthor{zhenjun zhou, chaowei jiang}

\def\corrEmail{zhouzhj7@mail.sysu.edu.cn, chaowei@hit.edu.cn}

\begin{document}
\onecolumn
\firstpage{1}

\title[Rotation Mechanism]{The Mechanism of Magnetic Flux Rope Rotation During Solar Eruption} 

\author[\firstAuthorLast ]{\Authors} 
\address{} 
\correspondance{} 

\extraAuth{}

\maketitle

\begin{abstract}
Solar eruptions often show the rotation of filaments, which is a manifestation of the rotation of erupting magnetic flux rope (MFR). Such a rotation of MFR can be induced by either the torque exerted by a background shear-field component (which is an external cause) or the relaxation of the magnetic twist of the MFR (an internal cause).
 For a given chirality of the erupting field, both the external and internal drivers cause the same rotation direction. Therefore, it remains elusive from direct observations which mechanism yields the dominant contribution to the rotation. In this paper, we exploit a full MHD simulation of solar eruption by tether-cutting magnetic reconnection to study the mechanism of MFR rotation. 
In the simulation, the MFR's height--rotation profile suggests that the force by the external shear-field component  is a dominant contributor to the rotation. Furthermore, the torque analysis confirms that 
it is also the only factor in driving the counterclockwise rotation. On the contrary, the Lorentz torque inside the MFR makes a negative effect on this counterclockwise rotation.

\tiny
 \keyFont{\section{Keywords:} magnetic fields, magnetohydrodynamics (MHD), methods: numerical, sun: corona, sun: flares}   
\end{abstract}

\section{INTRODUCTION} 
Solar eruptions are the most spectacular phenomena that occur in the solar corona, often (but not always) releasing coronal mass ejections (CMEs) into interplanetary space. A magnetic flux rope (MFR), characterized by a twisted and writhed topological structure of magnetic fields, is generally accepted as a basically magnetic configuration underlying the phenomenon of CMEs \citep{Vourlidas2013}.
A southward orientation of its front-side magnetic field may cause severe geoeffectiveness of the space environment \citep{Gosling1993,webb1994}. 
During its eruption and propagation, a rotational motion is also observed, which
modulates the orientation of the MFR \citep[][and references therein]{Gibson2006,Green2007,Zhou2020}.
Therefore, unveiling the physics that triggers the rotation of erupting MFR  is a key issue in understanding the space weather and its terrestrial effects.

Several theories of how the flux rope rotates are competing in the last decades. 
For an MFR existing prior to an eruption, when its twist exceeds a threshold, a helical kink instability \citep[KI;][]{Torok2004} may occur. Sigmoids observed in Soft-X-ray  and rotating filaments in EUV channels are interpreted to be outlining a MFR undergoing KI \citep{Gibson2006}. Observationally, the rotation direction of a filament apex has a strong one-to-one relationship with the filament helicity (chirality): 
For positive/negative helicity, filaments (with sinistral/dextral chirality) rotate clockwise (CW)/counterclockwise (CCW) \citep{Green2007,Zhou2020}. 
Numerical simulations \citep{Kliem2012}  find this relationship to be
consistent with the conversion of twist into writhe under the ideal MHD constraint of helicity conservation, suggesting the rotation mechanism as the relaxation of tension in the twisted field. 
\citet{Isenberg2007} propose an alternative rotation mechanism:
a Lorentz force exerted on the MFR legs from the misalignment between the MFR's toroidal current and the external toroidal field drives the MFR rotating. It cannot be easily distinguished which mechanism is responsible for the rotation, because both of these two mechanisms writhe the flux rope axis in a similar manner from observations. For a given chirality, 
the rotation driven by the external shear-field component yields the same rotation direction as that driven by the helical kink instability\citep{Isenberg2007,Kliem2012}.

To determine which mechanism dominates during the flux rope rotation, \citet{Kliem2012} carried out  a comparative study of these two mechanisms: 
Both the forces by twist and shear are potentially significant contributors to the rotation.
Their contribution can be disentangled if the rotation and rise profiles in the observations are simultaneously compared with model calculations. 
The profile of MFR's rotation vs. height in the simulation shows that the twist-driven rotation tends to saturate in the low corona (a height range up to several times of the distance between the footpoint of the MFR),
while, the shear-driven rotation distributes across a larger height range.
For parameters characteristic of filament source regions,
the shear field is usually present and dominates the rotation in the corona even if the twist is sufficiently high to trigger the helical kink instability. For a considerable rotation angle (angles of order $90\degree$ and higher), this shear-driven component is required.

Recently, \citet{Jiang2021} performed a fully three-dimensional (3D) magnetohydrodynamic (MHD) simulation demonstrating a runaway tether-cutting reconnection alone initiates a solar eruption. 
In the simulation, a MFR with dextral chirality is built up with a reverse-S shape through reconnection of the sheared arcades.
While rising, the simulated MFR rotates CCW, which compares favorably with the observation rule of filament chirality and rotation direction \citep{Zhou2022}.

Lacking the regular observations of the coronal
magnetic field, we employ the same simulation result as presented in \citet{Zhou2022} as the
observation substitution to look into the structure evolution and force distribution of the MFR to
explore the underlying mechanism of the MFR rotation. In this direction, we also calculate the
torque associated with the Lorentz force.
The numerical model is described in Section~\ref{MHD}. A whole picture of the simulated eruption and
a comparative analysis of height--rotation profile with the result of \citet{Kliem2012} are
given in Section~\ref{observation}, and a systematic survey of the global Lorentz torque is given in Section~\ref{Force}. Discussion and summary are presented in Section~\ref{DIS}.

\section{MHD SIMULATION}\label{MHD}
\citet{Jiang2021} performed a high-accuracy, fully 3D MHD simulation that examined a fundamental mechanism for the initiation of solar eruptions. They found that a current sheet can be slowly formed within a bipolar magnetic field as driven by quasi-static shearing motion at the photosphere. Then once magnetic reconnection sets in at the current sheet, it triggers and drives the eruption impulsively. In this process, an MFR is built up from the sheared arcade during the eruption, similar to the classic tether-cutting model. In this paper we study the evolution of the erupting MFR by using a similar simulation run like the one in \citet{Jiang2021} but with a slightly lower resolution such that the evolution of the MFR can be followed longer. The exactly similar run is used by \citet{Zhou2022} to demonstrate the CCW rotational motion of the MFR.

The simulation solves the full set of MHD equations including both solar gravity and plasma pressure. The initial condition consists of a bipolar potential magnetic field and a hydrostatic plasma ($\rho=2.3\times 10^{-15}$ \si{\gram\per\centi\meter\cubed}) stratified by solar gravity with typical coronal temperature ($T=10^6 K$). To energize the coronal field, we applied shearing flows along the PIL, which are implemented by rotating the two magnetic polarities at the bottom surface in the same CCW direction. The surface flow is stopped once an eruption is triggered (the time is referred to as $t=0$ in the following). During the quasi-static evolution phase driven by the shearing motion, a current sheet is gradually built up. Because we used no explicit resistivity in the MHD model, magnetic reconnection is triggered when the current sheet is sufficiently thin with its width close to the grid resolution, owing to the implicit, grid-dependent numerical resistivity.
The whole size of the computational volume extends as (-32, -32, 0) $< (x, y, z) <$ (32, 32, 64) with length unit ($l_0$) of 11.5 Mm. The grid resolution is 360 km, lower than  the resolution ($\geq$ 180 km) in \citet{Jiang2021},
here the reconnection rate is higher since with a lower resolution the numerical diffusion is larger. But when the resolution is high enough, the experiments with different higher resolutions (from 180 km to 22.5 km) suggest that the reconnection rate does not change much, that is, it seems to be independent on the Lundquist number when the Lundquist number is sufficiently high ($\geq 10^5$).
All the other setting are the same as descried in \citet{Jiang2021}, 
inculding the initial and boundary conditions.
Readers are referred to \citet{Jiang2021} for more details of the simulation settings. In \citet{Jiang2021}, the simulation reaches very high resolutions and the plasmoid instability is triggered in the current sheet, after which the magnetic topology becomes extremely complicated in small scales along with the formation of a large-scale MFR. Such a complexity substantially complicates our analysis of the large-scale evolution associated with the erupting MFR. 
To avoid this complexity, as mentioned above, here we have used the slightly lower-resolution run such that the basic evolution of the MFR during the eruption is not changed compared to the high-resolution run, except that the small-scale complex structure will not arise.
Moreover, we can run the simulation longer and thus follow a longer evolution of MFR with a lower resolution.

\section{Observation Features of Simulated Eruption}\label{observation}
\begin{figure}[h!]
\begin{center}
\includegraphics[width=\linewidth]{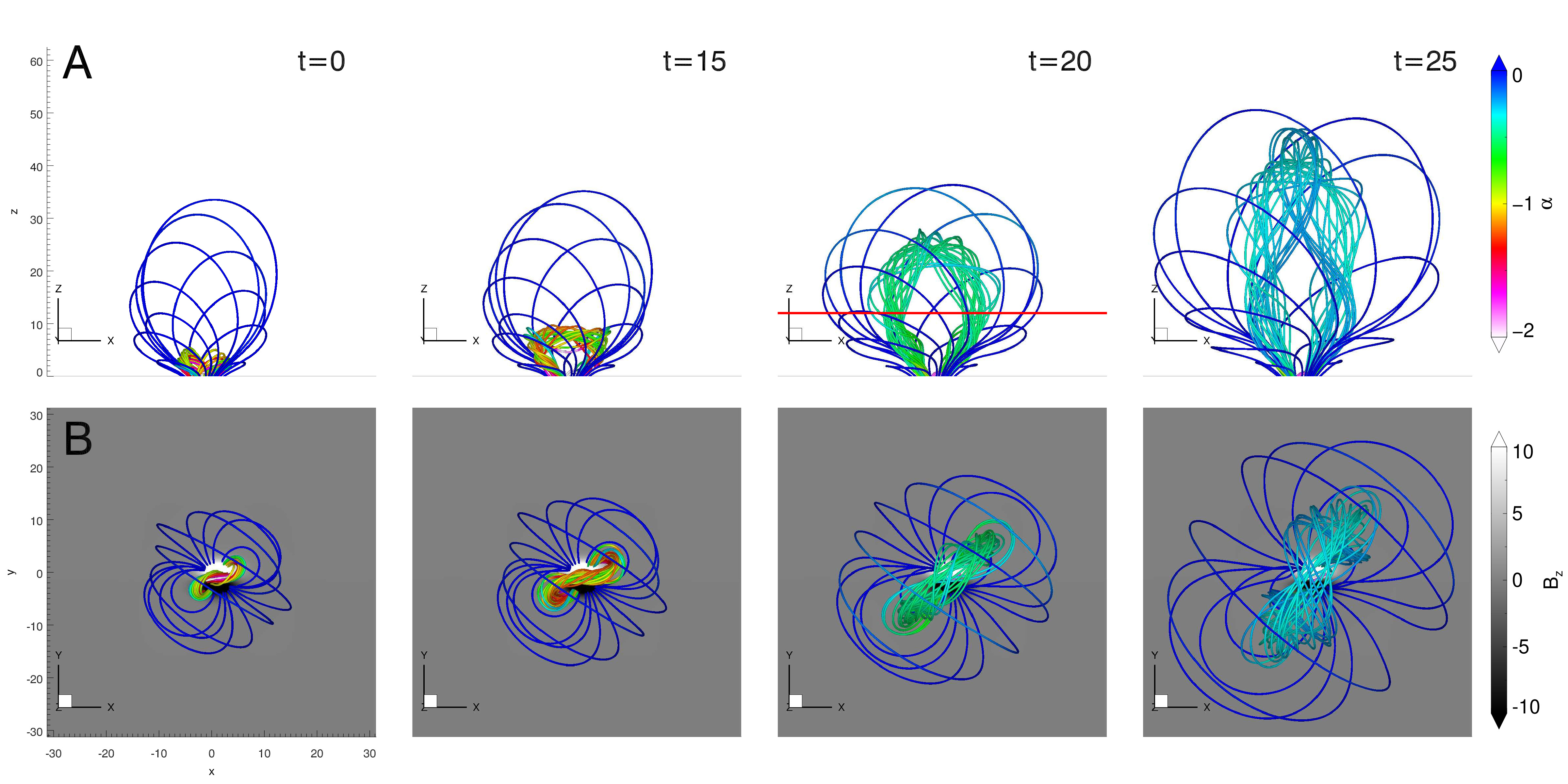}
\end{center}
\caption{Structural evolution of the eruption. \textbf{(A)} 3D perspective view of magnetic field lines colored by the force-free factor. Here the field lines are traced at fixed footpoints on the bottom surface, and they represent the core structure of the MFR and its ambient magnetic field. At $t=20t_s$, a red horizontal line gives the location of the slice in Figure~\ref{fig:3}(a). \textbf{(B)} Top view of the structure is shown in (a). The unit of time is 105 s. Also, see Supplementary Movie S1 for a high-cadence evolution of the eruption process.
}\label{fig:1}
\end{figure}

As shown in Figure~\ref{fig:1} (and accompanied Movie 1), the simulation demonstrates a typical solar eruption leading to a CME, as seen in many well-observed eruptions in different view angles \citep{Zhou2020,Zhou2022}. 
The core structure of evolving magnetic field is transformed from the pre-eruptive sheared arcades to twisted MFR that subsequently expands, and then transforms into a semi-circular shape stretching outward its overlying field,
finally displays a clear inverse $\gamma$ shape, indicative of a writhing motion of the apex (Figure~\ref{fig:1}(a)). From the top view (Figure~\ref{fig:1}(b)), as the reconnection kicks in, an inversed S-shaped MFR is formed on-the-fly. After that, the MFR's axis shows a significant CCW rotation, straightening, and even flipping in shape from reversed  S to forward S.

For the first time, this simulation reproduces the predictions for the relationship between the chirality of the erupting field, sigmoid shape, and projected MFR's apex rotation in the tether-cutting model \citep{Green2007}, i.e., for positive (negative) helicity the filament apex rotates CW (CCW) when viewed from above, and both the filament and related sigmoid take on a forward (reverse) S shape. This is consistent with that from observations \citep{Zhou2020,Zhou2022}.

Through this CCW rotation, the original inverse-S-shaped MFR is straightened to a straight loop, implying a reduction of writhe \citep{Torok2010,Liu2012}. Furtherly, the initial increase of the flux rope twist by the transformation of writhe helicity is directly confirmed based on this simulation \citep{Zhou2022}, which contradicts the expectation of the kink instability since this instability transforms twist into writhe. 
So instead we turn to an alternative rotation mechanism that is controlled by  an external shear-field component\citep{Isenberg2007}. 
Both of them yield the same rotation direction for a given chirality, 
still, we can tell which rotation mechanism dominates by comparing both the observed height--rotation profile with the corresponding curves from models \citep{Kliem2012}.

\begin{figure}[h!]
\begin{center}
\includegraphics[width=\linewidth]{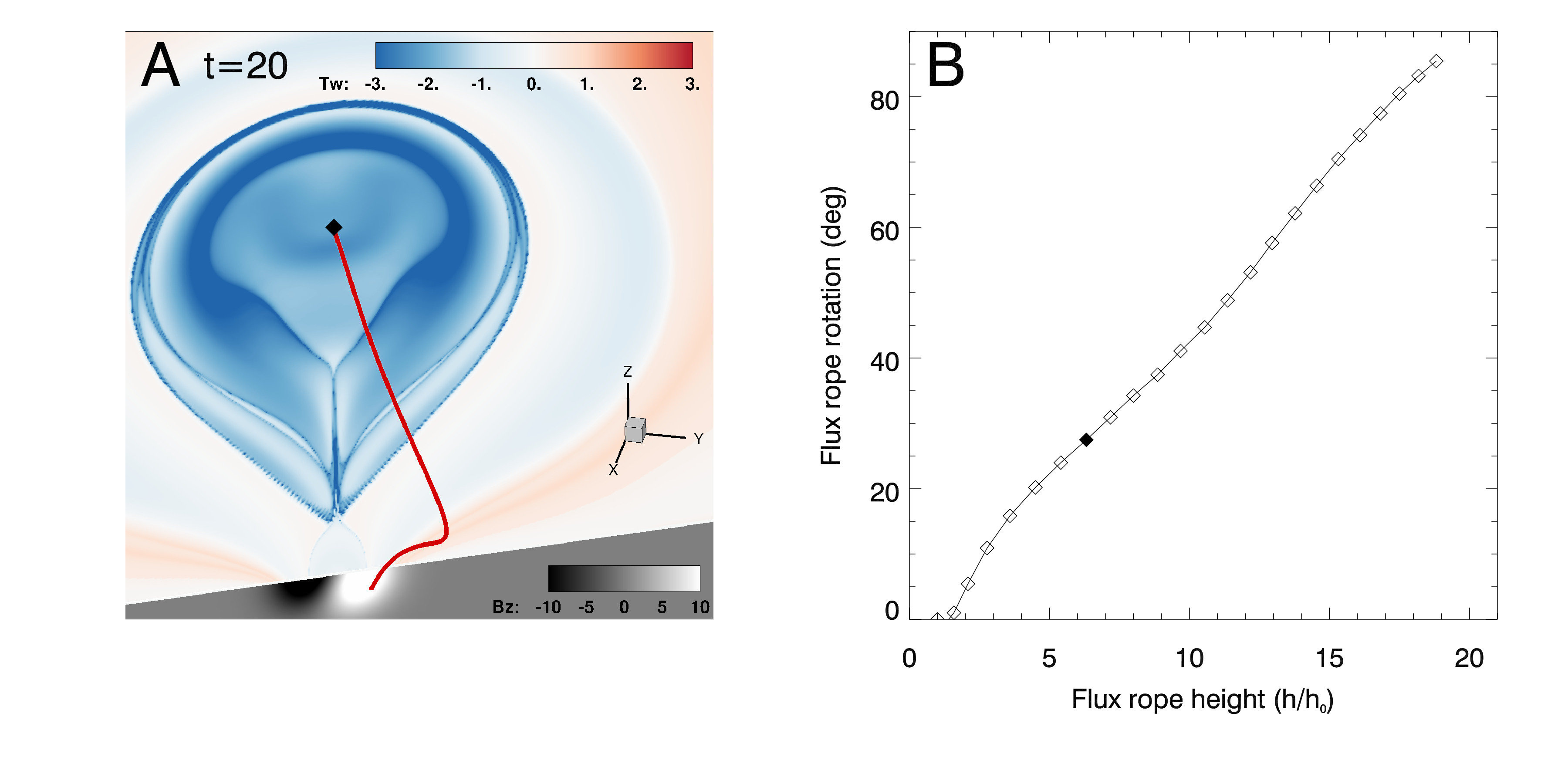}
\end{center}
\caption{
\textbf{(A)} The distribution of twist number (at $t=20t_s$) on the normal plane that is perpendicular to the tangent vector of MFR's axis (the red curve) at the apex point (black filled diamond, $h=6.32h_0$). The bottom plane (z=0) shows the Bz field in greyscale. 
\textbf{(B)} Comparison of flux rope rotation as a function of normalized apex height. The black-filled diamond marks out  their relation at $t=20t_s$.
}\label{fig:2}
\end{figure}

A powerful tool - the height--rotation profile proposed by \citet{Kliem2012} - can be used to diagnose the rotation mechanism of this simulated eruption. 
Here we use the rotation of MFR's axis as a proxy of MFR's rotation.
The determination of this modeled MFR's axis can be referred to \citet{Zhou2022}.
The rotation angle ($\Phi$) is measured as
the difference in angle between the tangent vector at the axis top, and its initial direction at $t_0$,
while the apex height ($h$) of MFR's axis is normalized by the initial apex height ($h_0=2.75 l_0$) of the axis (Figure~\ref{fig:2}(a)). 

The profile of rotation vs. height then can be obtained in this simulated solar eruption (Figure~\ref{fig:2}(b)). At the beginning (t=12),  the MFR is formed with its axis apex at the initial height ($h_0$) and the rotation angle is $0\degree$. 
A distinct monotonic increase in the rotation angle could be found as the MFR rise up. When it reaches the height of 19$h_0$, the axis rotates 85$\degree$. Apparently, there is no saturation of the rotation under the height of 19 $h_0$.
The parametric study by \citet{Kliem2012} shows that twist-driven rotation alone tends to saturate at lower heights ($<10h_0$, see Figure 6 in \citet{Kliem2012}) than shear-driven rotation. 
For parameters characteristic of the source regions of solar eruptions, if the rotation reaches angles of order 90$\degree$ and higher, the shear field must contribute the major part of the total rotation. In our previous study \citep{Zhou2022}, by calculating the twist of this modeled MFR during its eruption, we found that the twist of field lines winding around the rope axis increases, while the writhe of the rope axis decreases, which is distinct from kink instability (or a reverse process of kink-driving evolution). Thus, we concluded that the rotation of the rope is not twist-driven.

\section{force analysis}\label{Force}
The height--rotation profile implies that the force by an external shear-field component makes a significant contributor to the rotation.
This mechanism can easily be understood in a simplified scenario: a flux rope is settled in an ambient field. 
The misalignment between the flux rope's toroidal current and the shear field yields antisymmetric sideways Lorentz forces on opposite side legs, forcing the flux rope to rotate.
Gas pressure and gravity have a negligible effect on the magnetic structure in the corona. Thus, in this simulation, the only force to be considered is the Lorentz force $\vect F = \int \vect J\times \vect B dV$.

\begin{figure}[h!]
\begin{center}
\includegraphics[width=\linewidth]{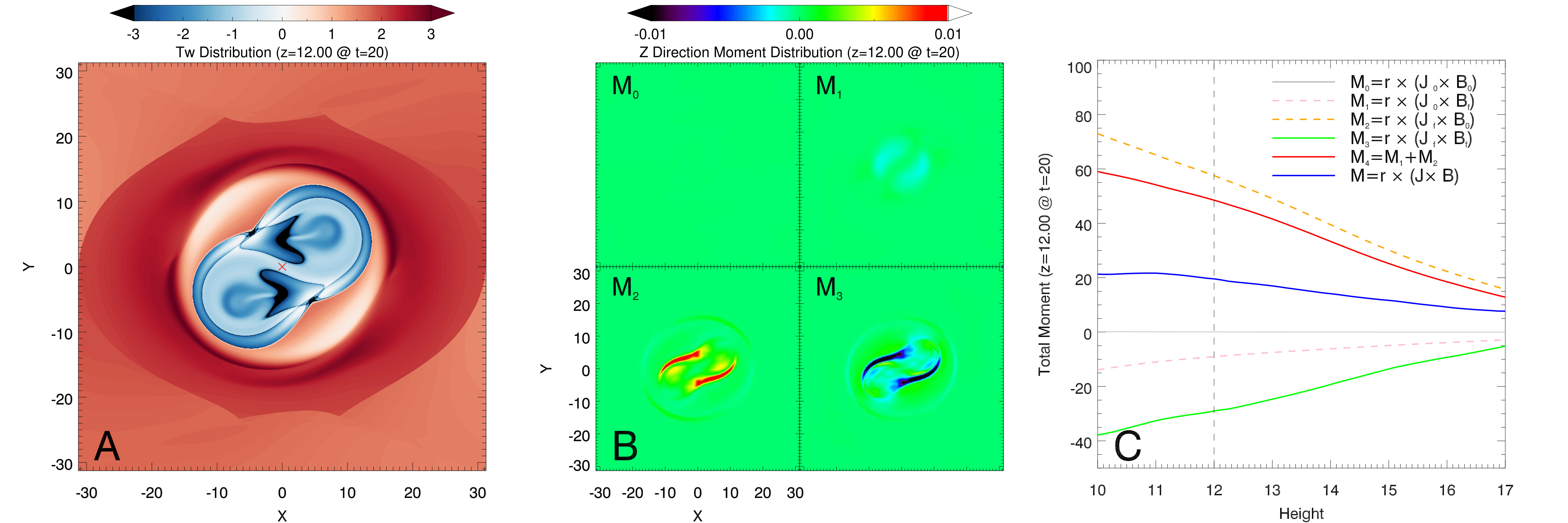}
\end{center}
\caption{Torque analysis at $t=20t_s$. \textbf{(A)} Distribution of twist number on the horizontal slice at z = 12 $l_0$.  The red cross symbol is the center of rotation. 
\textbf{(B)} Distributions of each z-direction moment component on the z-constant plane (z = 12 $l_0$).
\textbf{(C)} The z-direction component of Lorentz torque induced by the external field ($\vect M_0$, grey line), MFR itself ($\vect M_3$, green line),  their interaction force ($\vect M_4$, red line), and the total force ($\vect M$, blue line) integrated on each horizontal slice. The black vertical dashed lines show the location of the left panel. }\label{fig:3}
\end{figure}

As all the structures are centrosymmetric with respect to the origin on each horizontal plane ($z=z_0$),
the MFR and background can be distinguished from the twist distribution (Figure~\ref{fig:3}(A)) and the vertical ($\vect{z}$) component of the Lorentz torque  $\vect M = \int \vect r \times (\vect J\times \vect B) dV$ drives the structures on the horizontal plane to rotate. The positive/negative value in $\vect{z}$ direction means a CCW/CW rotation. In the simulation, the magnetic field $\vect{B}$ and current density $\vect{J}$ consist of those of both MFR and its background field.  
We assume that the background field $\vect B_0$ is a force-free magnetic arcade before the formation of the flux rope, which is thus taken as the pre-eruption field at t=0 since the flux rope only begin to form after the eruption starts. 
Since the force-free field $\vect B_0$ is
determined by the magnetic field at the bottom boundary (corresponding to the photospheric field),
which remains almost unchanged during the fast eruption,
we further assume that the background field $\vect B_0$ does not change with time. 
Then, to isolate shear-driven Lorentz torque from the total Lorentz torque, the magnetic field and current at time $t$ can be instantly decomposed into  $\vect B = \vect B_0 + \vect B_f$ and $\vect J = \vect J_0 + \vect J_f$ 
where $\vect B_f$ and $\vect{J_f}$ represent the
MFR’s magnetic field and electric current respectively. $\vect{J_0}$ is current associated
with the background magnetic field $\vect B_0$ at the initial time ($t=0$) when the MFR has not yet
formed, therefore it is considered to be constant with time.
The original data from the simulation has been remapped to a uniform grid, then the total ``Lorentz torque density'' $\vect M$ integrated on a z-constant plane can be decomposed as:
\[
\begin{split}
\vect M
={}& \sum_{x,y} \vect r \times (\vect J\times \vect B)
\\
={}& \sum_{x,y} \vect r \times ((\vect J_0 + \vect J_f) \times (\vect B_0 + \vect B_f))
\\
={}& \sum_{x,y} (\vect r \times (\vect J_0 \times \vect B_0)+ \vect r \times (\vect J_0 \times \vect B_f)+\vect r \times (\vect J_f \times \vect B_0) +\vect r \times (\vect J_f \times \vect B_f))
\\
={}& M_0+M_1+M_2 +M_3
\end{split}
\]
Of which, the components of ``Lorentz torque density'' consist of the background ($\vect M_0$), MFR itself ($\vect M_3$), and their interaction ($\vect M_1$ and $\vect M_2$). In particular, $\vect M_2$ refers to the torque induced by the current of the MFR with the background field, while $\vect M_1$ refers to that induced by the background current with the magnetic field of the MFR. 
Lorentz torque is the volume integration of the ``Lorentz torque density'' ($=\int \vect M dV=\vect M \overline{V}$). Due to the uniform grid,
the $\overline{V}$ is a constant value, no matter which horizontal slice is. Then for this comparative analysis, the integrated value of ``Lorentz torque density'' $\vect M$ can be used as proxy data for Lorentz torque. For simplicity, this so-called Lorentz torque $\vect M$ is used in the following. 
As an example, the distributions of each component of M at t=20 are demonstrated in Figure~\ref{fig:3}(B).
Along the height of 10-17 $l_0$, their relative contributions to the rotation then can be directly calculated. In this range, the horizontal slice can pass through the MFR's legs and ensure that the background field is not derived from the initial shear arcades.

A comparison between the contributions of these components is shown in Figure~\ref{fig:3}(C). Of which, the blue profile shows that the total Lorentz torque ($\vect M$) is always positive along the height, consistent with the observed CCW rotation. The Lorentz torque ($\vect M_2$, the orange dash line in Figure~\ref{fig:3}(C)), caused by the misalignment of the MFR-carried current and the sheared ambient field,
is the dominant factor for the rotation, not only it has a greater influence (with a larger absolute value than the others), but also it is the only factor (with a positive value) that drives a CCW rotation.
Compared to the other components in Figure~\ref{fig:3}(C), the $\vect M_0$ part is almost zero (its mean value $\overline{\vect M_0}=0.044$), because the background field is in a force-free equilibrium (i.e., $\vect J_0 \times \vect B_0 =0$) at the initial time.
It is worth noting that the Lorentz torque ($\vect M_3$) exerted by MFR's inner field component is negative, generating an adverse effect on the MFR's CCW rotation, and so does the $\vect M_1$.

\section{DISCUSSION AND SUMMARY}\label{DIS}
With the aid of \citet{Jiang2021}'s simulation, we investigated the rotation mechanism of erupting MFR formed in the runaway tether-cutting reconnection model.
The relationship between chirality, sigmoid shape, and apex rotation in this simulation is in line with that in observations. As the MFR ascends during the eruption, its apex rotates CCW.
The rotation angle of simulated MFR monotonically increases to 85$\degree$ and shows no sign of saturation at a relatively large  height $\sim 20h_0$.
\citet{Kliem2012} suggested that the rotation triggered by KI reaches saturation rapidly at a height comparable to the footpoint distance, corresponding to the degree of rotation to level off a much smaller height (a few $h_0$) above the photosphere. The rotation driven by the shear field acts over a greater height range.
Moreover, if the MFR rotates by large angles of order $90\degree$ and more, the shear field is found to be 
required and acts quite efficiently.

Further on, the force analysis confirms the driving force of the rotation originates solely from external sources.
The decomposition of the $\vect{z}$ direction component of Lorentz torque shows clearly that the part induced by the MFR's current and the background field,  $\vect M_2 = \int \vect r \times (\vect J_f \times \vect B_0)dV$, is the only positive component to drive the MFR  rotating CCW.
This sideways Lorentz force continuously adjusts the loop current's orientation to align with the external shear field. On the contrary, the  $\vect{z}$ direction component of inner Lorentz torque, $\vect M_3 = \int \vect r \times (\vect J_f \times \vect B_f)dV$, has a negative value, hindering the MFR's CCW rotation. 
Therefore, the strength of the external shear field serves as the dominant factor determining the CCW rotation.
Previous numerical simulations \citep{Kliem2012} suggest the KI yields the same rotation direction for the given chirality of the erupting field.
Our simulation has indicated the occurrence of the chirality-corresponding rotation doesn't necessarily reflect a relaxation of tension in the twisted field, and it quite suggests the opposite process: an enhancement of tension in the twisted field.
This has also been validated by 
our recent study \citep{Zhou2022}: a quantitative measurement in this simulation clearly shows conversion of writhe into twist during rotation.

It is worth noting that our research is based on the MHD simulation of CME initiation by tether-cutting reconnection in a single magnetic arcade  rather than a pre-existing unstable MFR. Therefore, although the inference from this model is consistent with the observed sigmoid--chirality relationship, it does not rule out other possibilities, because the initial magnetic topology being that of a sheared arcade is still intensely debated \citep{Patsourakos2020}.
Moreover, there also exist some other mechanisms, e.g., magnetic reconnection of rope field lines with the ambient field \citep{Jacobs2009,Cohen2010,Thompson2011,Vourlidas2011}, asymmetric deflection of the rising flux rope induced by adjacent coronal holes \citep{Panasenco2011} and/or local alignment with the heliospheric current sheet \citep{Yurchyshyn2008}, can cause the rotation of a MFR.   
All these theories are needed to be tested with a measurement of the coronal magnetic field in future works, to understand fully the mechanism of filament rotation during solar eruptions.

\section*{Data Availability Statement}
The original contributions presented in the study are included in the article/Supplementary Material, further inquiries can be directed to the corresponding author.

\section*{Author Contributions}
Z.Z. leads this work, C.J. performs the numerical simulations  and all contribute to the study.

\section*{Acknowledgments}
The authors wish to express their special thanks to the referee for suggestions and comments which led to the improvement of the paper.
This work is supported by the B-type Strategic Priority Program XDB41000000 funded by the Chinese
Academy of Sciences. The authors also acknowledge support from the National Natural Science Foundation of China (NSFC 42004142,42274203), Open Research Program of CAS Key
Laboratory of Geospace Environment, Science and Technology Project 202102021019 in Guangzhou, the Fundamental Research Funds for the Central Universities (grant No.HIT.BRETIV.201901), and Shenzhen Technology Project JCYJ20190806142609035.

\bibliographystyle{Frontiers-Harvard}
\bibliography{bibfile}


\section*{Figure captions}


\end{document}